\documentstyle[sprocl,epsfig]{article}

\def\be{\begin{equation}}
\def\ee{\end{equation}}
\def\bea{\begin{eqnarray}}
\def\eea{\end{eqnarray}}

\def\as{\alpha_{\mbox{\tiny S}}}

\def\mR{\mu_{\mbox{\tiny R}}}

\begin{document}
\begin{flushright}
HD-THEP-01-7\\
hep-ph/0102132
\\
\vspace*{1cm}
\end{flushright}                                                                

\title{QCD and Hadronic Final States at the LHC \footnote{Talk 
presented at the XXXth International Symposium on Multiparticle 
Dynamics, Tihany, Hungary, October 2000}}

\author{Carlo Ewerz}

\address{Institut f\"ur Theoretische Physik, Universit\"at Heidelberg,\\
Philosophenweg 16, D-69120 Heidelberg\\
E-mail: carlo@thphys.uni-heidelberg.de} 

\maketitle\abstracts{
Hadronic final states at the LHC will be an
interesting testing ground for QCD. A good
understanding of QCD radiation will also be
important for the discovery of new physics
at the LHC. I discuss some aspects of 
this subject and give a few examples.  
}

\section{Introduction}
The Large Hadron Collider will offer 
the chance of discovering the Higgs boson, 
supersymmetry, extra dimensions, or even more 
surprising new physics\footnote{Many of the remarks in this talk 
also apply to the upgraded Tevatron.}. 
The discovery of any new particle and even more 
so the precise determination of its properties will 
require a good understanding of the relevant background 
processes. The background processes as well 
as basically all production cross sections at the LHC involve 
strong interaction physics --- at least via the 
non--perturbative parton distribution functions (pdfs)
for the initial state hadrons. A good knowledge of 
QCD will thus be mandatory for any discovery at the LHC. 
At the same time the LHC will allow us to test QCD in a 
new kinematical region. This will make it possible to  
learn more about the complicated dynamics of 
QCD. 

Along with the high luminosity the accessible 
kinematical range is one of the most important features of the 
LHC. In parton collisions at the LHC Bjorken-$x$ 
will range from $1$ down to values as small as $10^{-6}$. 
This range of $x$ is comparable to the one 
studied at HERA but the corresponding values 
of $Q^2$ will be significantly higher. 
Due to this there are for example excellent prospects 
for obtaining improved parton distribution functions 
and for testing DGLAP evolution.  
In particular it will be possible to constrain the gluon 
distribution function from 
jet and photon production data. 
The kinematical range of the LHC will also allow one 
to study the interesting dynamics of small-$x$ QCD at larger 
momentum scales $Q^2$ at which a perturbative treatment 
becomes possible. 

A variety of aspects of QCD can be studied by 
measuring different properties 
of the hadronic final states emerging from proton--proton 
collisions. These final state properties include 
jet shapes, angular distributions, multiplicities, 
heavy quark fractions, fragmentation etc. 
These observables will be especially useful in investigating 
the challenging problems related to the hadronisation 
process and the transition region from hard (perturbative) to 
soft (non--perturbative) interactions in QCD. 

In the following I will briefly discuss three examples 
that highlight some of the aspects mentioned 
above.\footnote{The choice is or course biased 
by my own interests and prejudices.} 
For a much more detailed overview of issues 
related to studying QCD at the LHC and a complete 
list of references the reader is referred to \cite{Cernrep}. 

\boldmath
\section{Small-$x$ Dynamics in Forward Dijets and their 
Azimuthal Decorrelation}
\unboldmath
The dynamics of QCD in the high--energy limit 
(or correspondingly at small Bjorken-$x$) is 
expected to exhibit very interesting properties. 
Processes in which the squared energy $s$ 
is much larger than the momentum transfer $t$, 
$s \gg t \gg \Lambda^2_{\mbox{\scriptsize QCD}}$,  
are typically enhanced due to the existence of 
large $\ln(s/t)$ logarithms. 
These logarithms can be resummed to all orders 
in perturbation theory, resulting in the celebrated BFKL 
equation \cite{FKL,BL}. The corresponding cross sections 
grow like powers of the energy, $\sigma \sim s^\lambda$, 
with $\lambda= \as 12 \ln 2/\pi \simeq 0.5$. 
However, in many processes it has turned out to be difficult to 
disentangle BFKL physics from DGLAP evolution. 

\begin{figure}[ht] 
\begin{center}
\mbox{\epsfig{figure=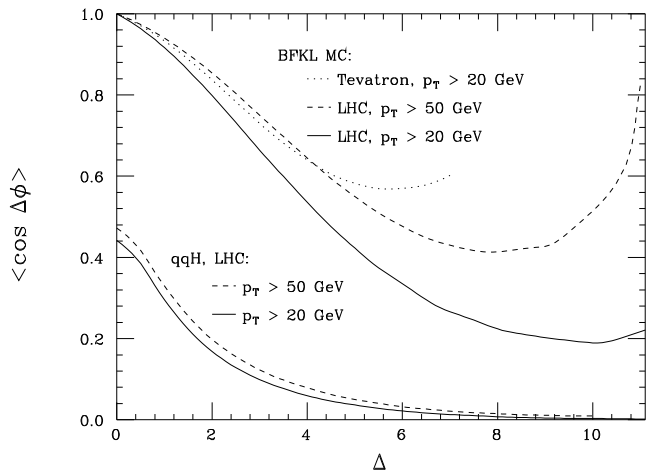,width=10.5cm}}
\end{center}
\caption{The decorrelation in azimuthal angle in dijet production at the 
Tevatron ($\sqrt{s}=1.8$~TeV) and LHC ($\sqrt{s}=14$~TeV)
as a function of dijet rapidity difference $\Delta y$.  
The upper curves are: (i) Tevatron, $p_T>20$~GeV (dotted curve),
(ii) LHC, $p_T>20$~GeV (solid curve), and (iii) LHC, $p_T>50$~GeV
(dashed curve).  For comparison, 
the lower curves are for dijet production in the process
$qq\to qqH$. (Figure from \protect\cite{Orr}.)
}\label{fig_bfklphi}
\end{figure}

The production of jet pairs with large rapidity separation $\Delta y$ 
in hadron collisions has been suggested \cite{MN} 
as a particularly well--suited process for isolating 
small-$x$ effects. On the parton--level the corresponding 
cross section is predicted to rise as 
$\hat{\sigma}_{jj} \sim \exp(\Delta y)$. 
In order to compare with the experimentally measured 
cross section, however, this subprocess has to be convoluted 
with the pdfs of the colliding hadrons. Unfortunately, the 
pdfs decrease faster with $\Delta y$ than $\hat{\sigma}_{jj} $ 
increases, and the small-$x$ effects are again very 
difficult to observe. 
It has been pointed out \cite{Vittorio,James} 
that the decorrelation 
in the relative azimuthal angle $\Delta \phi$ 
of the two jets is relatively insensitive to the pdfs and in 
fact should lead to a clearly visible effect of BFKL 
dynamics. Here $\Delta \phi$ is defined as 
$\vert \phi_1-\phi_2 \vert - \pi$. In lowest order 
the two jets originate from the scattering of 
two partons via one--gluon exchange and are thus 
produced back--to--back, i.\,e.\ $\Delta \phi=0$. 
The BFKL  calculation of this quantity predicts a 
much larger decorrelation than is expected at 
fixed order due to the emission of 
additional gluons between the jets. 
This effect is illustrated in Fig.\ \ref{fig_bfklphi}. 
The results in this figure have been obtained \cite{Orr} using a 
Monte Carlo method to evaluate the leading order BFKL 
equation. This approach leads to a more realistic 
prediction than the analytic solution because it allows one 
to overcome some deficiencies inherent in the latter by 
implementing for example a running coupling and 
correct kinematical constraints for the produced gluons. 
The curves are obtained for different lower cuts on the 
transverse momenta of the jets, $p_{T1},p_{T2}>p_T$, 
and for LHC as well as for the Tevatron. 
The characteristic BFKL decorrelation is clearly 
visible. It increases with increasing $\Delta y$ before 
flattening off and finally decreasing again at the 
kinematical limit. 
Recent data obtained by the D0 collaboration \cite{D0} 
are in reasonable agreement with the predictions. The 
effect should be even more pronounced at the LHC due 
to the larger energy that also makes a larger range in 
$\Delta y$ available. 

\section{Minijet Multiplicities in Higgs Production}
The discovery of the Higgs boson at the LHC will obviously 
require a good prediction for background processes 
leading to the same final state as a decaying Higgs 
boson. But its identification could also be affected by 
a very large number of mini--jets produced in the 
same hard scattering process. The number of 
mini--jets in a hard collision is therefore an 
important aspect of the hadronic final state. At the 
same time it is an interesting observable for studying 
the dynamics of QCD radiation in a hard scattering 
process. 

A mini--jet is a jet with a transverse momentum 
above some resolution scale $\mR$ which is much smaller 
than the hard scattering scale $Q$. The mini--jet rate 
at small $x$ involves not only large logarithms of $1/x$ 
but also additional large logarithms of the form $T=\ln(Q^2/\mR^2)$ 
that need to be resummed. 
The mean number of mini--jets in a hard small--$x$ process has 
been computed \cite{minijets1,minijets2} 
based on the BFKL formalism. The results 
are expected to hold also in the framework of CCFM evolution 
based on angular ordering of gluon emissions. The results include 
all terms of the form $(\as \ln x)^n T^m$ with $1\le m\le n$. 
The terms involving $m=n$ are called double--logarithmic (DL), 
whereas the terms with $m<n$ correspond to single--logarithmic (SL) 
corrections. 

\begin{figure}[ht]
\begin{center}
\input{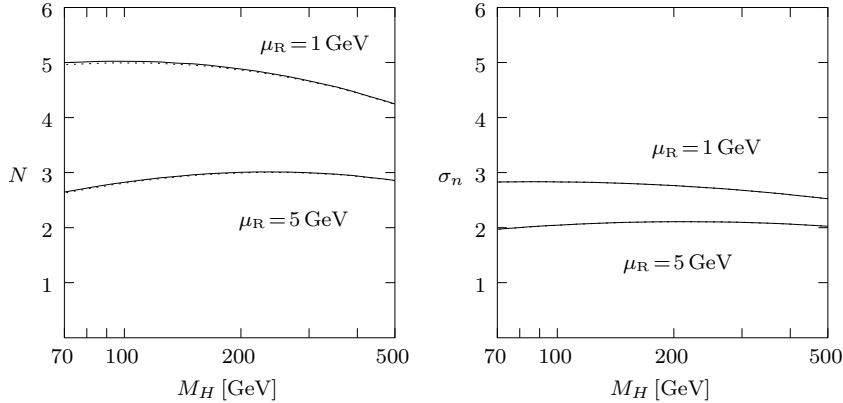}
\end{center}
\caption{The mean value and dispersion of the number of (mini-)jets
    in central Higgs production 
    at LHC for two different resolution scales $\mR$. Solid lines 
    show the SL results up to the 15th order in perturbation theory, 
    dashed lines correspond to the DL approximation.}\label{fig_higgsjets}
\end{figure}

The central production of a Higgs boson is a typical example 
for the application of the formalism developped in 
\cite{minijets1,minijets2}. The dominant production mechanism 
is expected to be gluon--gluon fusion, and the momentum 
fractions of the gluons $x=M_H/\sqrt{s}$ are of order $\sim 10^{-3}$. 
Fig.\ \ref{fig_higgsjets} shows the mean number $N$ of mini--jets 
and its dispersion $\sigma_N$ as a function of the Higgs mass. 
In these numbers we do not include the jets originating from the 
proton remnants. 
\label{Hg}
The DL terms approximate the result very well and 
the SL terms are less significant in this case. In total 
the mini--jet multiplicity does not vary much with the 
Higgs mass. 
Even for low resolution scales $\mR$ the number of mini--jets is 
fairly low, and the identification of the Higgs boson should 
not be seriously affected. 

\section{Jet Shapes in Hadron Collisions}
Very interesting properties of the hadronic final states 
are encoded in the jet shape variables like thrust, $C$-parameter, 
etc. A variety of these infrared and collinear 
safe observables has been studied in $e^+e^-$ collisions and in DIS. 
It has been found that they exhibit significant non--perturbative 
power corrections. Jet shapes are 
therefore optimal observables for studying the interplay 
between perturbative and non--perturbative effects.  
The mean value of a given event shape variable $\cal{F}$ has the form 
$ \langle {\cal F} \rangle = 
  \langle {\cal F}_{\mbox{\scriptsize pert}} \rangle + 
  \langle {\cal F}_{\mbox{\scriptsize NP}} \rangle $\,.
The non--perturbative correction to the perturbative result 
$\langle {\cal F}_{\mbox{\scriptsize pert}}\rangle$ is power--suppressed, 
$\langle {\cal F}_{\mbox{\scriptsize NP}} \rangle 
= C_{\cal F} \mu/Q^p\,,$
where 
the exponent $p$ can be obtained via a renormalon analysis \cite{Martin}. 
Remarkably, the power corrections to different observables 
have been found to be (approximately) universal, i.\,e.\ the scale 
$\mu$ is the same for certain classes of event shapes, 
and ${\cal C_F}$ is a perturbatively calculable coefficient 
depending on the variable under consideration. 
This universality holds to within $\sim 15 \%$ which is quite 
surprising for a quantity that {\em a priori} does not need 
to be universal at all. 
The dispersive approach to power corrections \cite{BPY} goes a 
step further and assumes that the notion of an (effective) strong 
coupling constant $\alpha_{\mbox{\scriptsize eff}}$ can be 
extended to very low momenta in the sense that its integral moments 
have a universal meaning. Then the magnitude of the non--perturbative 
power correction (i.\,e.\ $\mu$) can be related to low--momentum 
averages of the coupling, for example to 
\be
  \alpha_0(2\,\mbox{GeV})= \frac{1}{2\,\mbox{GeV}} 
  \int_0^{2\,\mbox{\scriptsize GeV}}  \,
  \alpha_{\mbox{\scriptsize eff}}(q) \, {\rm d}q\,.
\ee
The value for this quantity as extracted from 
$e^+e^-$ and in DIS data is around $0.5$. 
For a recent review of the 
phenomenology of power corrections see \cite{Yuri}. 

Power corrections are known to originate from 
non--perturbative effects in the hadronisation 
process. It would certainly be very interesting 
to confirm the universality of power corrections 
also in in the environment of hadronic collisions 
at the LHC, or alternatively to identify 
characteristic differences to $e^+e^-$ collisions 
and DIS. The theoretical description of 
power corrections in hadronic collisions is 
more complicated than in those cases, and 
so far there have been only very few 
theoretical studies \cite{Mike}. 
Additional difficulty originates from 
gluon radition from 
the initial state particles before a hard interaction 
takes place. This gluon radiation can change the 
geometry and thus affect the event shape variables. 
Another potential difficulty is related to the definition of the 
hard scale $Q$ to which the power suppression 
refers. In hadronic 
collisions there is no hard (perturbative) scale 
in the initial state (like the center--of--mass 
energy in $e^+e^-$ collisions or the photon virtuality 
in DIS). Thus the hard scale and the hemispheres 
relevant for the theoretical analysis need to 
be defined using the hard {\em final state} particles 
which in turn can only be observed as jets. This probably 
requires a very careful treatment of the details of the 
jet definition in use. 
Despite these theoretical obstacles I expect  
the investigation of power corrections to event 
shape variables to become a very interesting 
and important class of measurements at the LHC. 

\section{Conclusions}
The LHC will offer ample opportunity to test 
and to extend our understanding of many aspects 
of QCD. 
At the same time a good knowledge of QCD will 
be essential for fully exploiting the discovery potential 
of the LHC. 

\section*{Acknowledgments}
It is a pleasure to thank Bryan Webber for fruitful collaboration 
and Yuri Dokshitzer and James Stirling for interesting discussions.

\end{document}